# THE IMPACT OF DISINFORMATION ON A CONTROVERSIAL DEBATE ON SOCIAL MEDIA

**Salvatore Vilella**\*
Department of Computer Science
University of Turin
Turin, Italy
salvatore.vilella@unito.it

**Alfonso Semeraro**
Department of Computer Science
University of Turin
Turin, Italy
alfonso.semeraro@unito.it

**Daniela Paolotti**
ISI Foundation
Turin, Italy
daniela.paolotti@isi.it

**Giancarlo Ruffo**
Department of Computer Science
University of Turin
Turin, Italy
giancarlo.ruffo@unito.it

June 30, 2021

## ABSTRACT

In this work we study how pervasive is the presence of disinformation in the Italian debate around immigration on Twitter and the role of automated accounts in the diffusion of such content. By characterising the Twitter users with an *Untrustworthiness* score, that tells us how frequently they engage with disinformation content, we are able to see that such bad information consumption habits are not equally distributed across the users; adopting a network analysis approach, we can identify communities characterised by a very high presence of users that frequently share content from unreliable news sources. Within this context, social bots tend to inject in the network more malicious content, that often remains confined in a limited number of clusters; instead, they target reliable content in order to diversify their reach. The evidence we gather suggests that, at least in this particular case study, there is a strong interplay between social bots and users engaging with unreliable content, influencing the diffusion of the latter across the network.

## 1 Introduction

Disinformation episodes have been well documented much before the invention of the Internet, with the term 'fake-news' being commonly used from the end of the 19th century (1). Even if these phenomena are nothing new, these issues have been recently taken into serious consideration at both scientific and political levels, so that many national and supranational institutions have considered the related technical and ethical problems. For example, the Council of Europe has released a report in 2017 (2) pointing out that information pollution (and its variants such as mis-information, dis-information, and mal-information) is seen as a serious risk for democracies, and its role into manipulating the public discourse is a problem worth investigating: we need to understand, among the other things, if it can be linked to a declining trust into evidence based journalism, to the growing popularity of news outlets that promote low-quality information, and to a measurable impact of computer assisted devices, such as bots, that contribute significantly to a self-feeding diffusion process that could amplify the virality of fabricated content. This has serious implications on public discussions regarding different topics, from public health (see the so called infodemic phenomenon that characterised much information regarding COVID-19) to climate change, from economics to - quite naturally - politics.

---

\*Corresponding author

In the present work we study how frequently content from *untrustworthy* media outlets is shared in the particular study case of the Italian Twitter debate around migrants; we argue that this is a relevant case study because it has been quite controversial and heavy on Italian politics in the last years, considering that the contrast to illegal immigration is still one of the key points in the right-wing political parties' agendas (3). Therefore, we expect that immigration will be at the centre of the next national electoral campaigns. We aim to evaluate the presence of social bots in such a debate, to assess how bots' behaviour is connected with disinformation; particularly, we investigate if and how the interplay between bots and the spread of disinformation have an impact on the virality of a single piece of news.

The impact of disinformation and fake news on social media debates has been studied extensively during the last years. One of the most common ways of selecting the sources is by hand-picking a selection of news media outlets that are well-known disinformation spreaders. This selection is usually done by referring to one or more of the many existing services that monitor the quality of information and debunk viral fake news. This or similar approaches are followed by many (4; 5; 6; 7) - including us, as detailed in Section 2.3 - and it finds a crystal clear explanation in the seminal work by Lazer et al. (8), where the authors state that they "*advocate focusing on the original sources — the publishers — rather than individual stories, because we view the defining element of fake news to be the intent and processes of the publisher*". Focusing on sources, rather than on individual fake stories is likely to be crucial, because it allows to expand from the very specific phenomenon of fake news to the much more complex and multi-faceted world of *information disorders*, a very general term that includes many different aspects such as fake-news, propaganda, lies, conspiracies, rumours, hoaxes, hyper-partisan content, falsehoods or manipulated media, that are usually generalised within three main categories (see (2), for instance, for a fine grained classification of information disorders phenomena): disinformation, misinformation, and malinformation, namely all those stories that might not be entirely true nor false, but are presented in a partial (and often malicious) way. In particular, our case study leans towards understanding disinformation, even though we are aware that applying a single label to such complex phenomena can be misleading.

We set up our experiment within the context of the Italian online debate on the topic of immigration, specifically on Twitter. The online debate around migrations has been studied recently in several national and cross-national contexts. Some scholars observed that the topic of migrations is often extremely polarised (9; 10) and it displays a very high level of "mediatization" (11), with high-level politicians and media outlets being involved in the discussion (12). Disinformation - i.e. maliciously spread news that are often partially or entirely fake - in such discussions can be instrumental in targeting both politicians (13) and migrants (13; 14) and the general attitude can depend on the particular country or on the events under examination (15).

In general, as this is a strongly politicised topic, it is not exempt from the effects that disinformation has on other similar topics. The prevalence of disinformation content in different online debates has been studied (5; 4; 16; 17). Specifically, in (16) the authors study a context very similar to ours, adopting a comparable approach also in the selection of disinformation content, finding evidence of connections between the Italian disinformation sources and other European counterparts, with the majority of such content being spread in the Italian conservative and far-right political environment.

Another important aspect that we explore is the presence of bots and their role in the spreading of malicious content. A *social bot* can be defined as "*a computer algorithm that automatically produces content and interacts with humans on social media, trying to emulate and possibly alter their behaviour*" (18). The impact of social bots has been assessed in many different contexts, following different approaches. Results are not always perfectly aligned, as they strongly depend on the particular study case; there is though general agreement on the fact that they influence - to different extents - the conversation and, particularly, that there is a strong interplay between bots and humans, with the latter being crucial in the virality of content (19; 20; 21). Bots can often give a negative contribution to the discussion (22), disrupting communications (23) or increasing polarisation (21). Still, their contribution is not always as evident. In other experimental settings, even though a clear presence of bots has been found, the impact on the conversation appeared limited (24; 25); the already cited work by Vosoughi et al. (19) gives a crucial contribution in understanding that the role of bots is often non-trivial, as well as deeply rooted into their interaction with human users. Finally, Shao et al. in (20) have produced a considerable amount of empirical evidence aiming at better understanding the role of social bots in the spread of low-credibility content. They found that social bots play a disproportionate role in spreading articles from low-credibility sources, amplifying the diffusion of such content in the early spreading moments, before an article goes viral. Most importantly, bots target users with many followers through replies and mentions, exploiting human vulnerabilities to this kind of manipulation: real users are fooled as they are likely to re-share content posted by bots. Hence, bots play a fundamental role in spreading disinformation, even if more influential users that are targeted and easily manipulated to re-share low-credibility posts should probably be blamed the most.

All the aspects mentioned in this introduction - disinformation, social bots, opinion fragmentation - are all ingredients that are typical of online debates about sensitive issues such as immigration, making it an ideal case study to answer some research questions. Therefore, we build this analysis with the purpose to extend significantly the preliminary analysis on the immigration Italian debate we previously made on Twitter (26), where we found a strongly clustered network reflecting a general opinion fragmentation on such a divisive topic; in fact, the related retweet network was composed of different communities, consisting of users clustered around famous politicians and news media with a specific stance towards immigration. In the present work, we aim to deepen the understanding of the characteristics of the debate around immigration on Twitter, and how it is influenced by disinformation dynamics and bots. Our work revolves around the following questions:

- how pervasive is the presence of disinformation in the Italian debate around immigration on Twitter? Can we define a measure to quantify the action of a user to share content from untrustworthy media outlets? Are there communities that are more affected by this behaviour?
- what is the impact of social bots? is there an interplay between the action of social bots and the spreading of untrustworthy content?
- What are the key factors that maximise the probability of success in terms of spreading of untrustworthy media content?

## 2 Methods

### 2.1 Datasets and Retweet Network

In this study we use two datasets:

- the TWITA dataset (27): it is an ongoing collection of tweets identified as being written in Italian; the collection comprises hundreds of millions of tweets, starting from February 2012, without any filter but the language. We used TWITA multiple times throughout the paper, to collect random samples of tweets to build null models and baselines for our analysis.
- the TWITIMM ("Tweets in Italian on Immigrants") dataset (26): it was created for our previous work of analysis on the Italian Twitter debate around immigration, and it is composed of almost 6 millions tweets in Italian, collected using Twitter's Stream APIs, and filtered on keywords **migranti, immigrati, immigrazione** (in English: migrants, immigration). TWITIMM spans from August 2018 to August 2019 - the year of the so-called "first Conte's Government", when the Prime Minister Giuseppe Conte was leading a right-wing majority for whom the fight against illegal immigration was at the top of the Government agenda: for this reason, immigration and migrants' rights where among the most debated topics in the Country. From TWITIMM we extracted a set $U$ of more than 200,000 unique users.

The study of how information spreads on the underlying digital networks composed by users' interactions has shown to be crucial to study information diffusion on social media (28), hence we decided to adopt a network analysis approach and to model the interactions between users accordingly. We are particularly interested into modelling the *retweets*, i.e. the act of sharing someone else's tweet, often seen as a form of endorsement (29; 30; 31). Therefore, from TWITIMM we build a retweet graph: $G = \{N, L\}$ whose nodes $s, t \in N$ are the Twitter users in our dataset, and directed links $l = (s, t) \in L$ are established when $s$ has retweeted one tweet created by $t$ at least once. Every link $l$ has a weight $w$ that represents how many times user $s$ retweeted a content created by user $t$. The resulting graph contains more than 200,000 users and 2 millions edges.

It has to be noticed that many other classes of networks can be built based on users' interactions on Twitter: for example, we can create links between users who reply one to the other or who are mentioned by another user (mention/reply networks), or we can "follow" the tweet itself to build its diffusion tree (cascade networks). However, as already raised above, in our study we are interested to analysing relationships that are more likely explained in terms of homophily or implicit/explicit endorsement; in fact, there is a considerable volume of empirical evidences that homophily characterises retweet networks with respect to mention networks (e.g., see (31)). Moreover, cascades are difficult to reconstruct due to limitations of the basic Twitter APIs, whereas retweets are easily provided and properly annotated to distinguish them from original tweets and to recover the author of the original post.

Given that we are able to estimate if a user is more likely to share content of dubious quality, or if it is a bot mimicking human behaviour (astroturfing) through means of several algorithms trained to identify such bots, we aim at discriminating virality patterns when they are controlled mainly by "un/reliable humans", or by "'un/reliable bots". We will introduce our Untrustworthiness measure and the BotScore respectively in Sections 2.3 and 2.4, so that we can first focus on the metrics that can be used to measure the virality of a message. To this end, we want to:

- count the total number of users that retweeted a content;

| Community ID | Size | Internal Link Density | Political area / Characterisation | Inferred Stance Towards Immigration |
|---|---|---|---|---|
| RT1 | 116,831 | $1.5 \cdot 10^{-3}$ | Left, Centre-left, Democrats | Positive |
| RT2 | 34,174 | $1.93 \cdot 10^{-2}$ | Right, Far-right, Hoaxers, News Media | Negative |
| RT3 | 27,845 | $2.4 \cdot 10^{-3}$ | *League* party, Right, News Media | Negative |
| RT4 | 9,553 | $3.5 \cdot 10^{-3}$ | 5 Stars Movement, News Media | Mixed |
| RT5 | 9,225 | $2.4 \cdot 10^{-2}$ | News Media, All News outlets | Neutral |

Table 1: Table describing the communities found in $G$ in terms of size, internal link density, political leaning and/or general characterisation and inferred stance towards immigration (26).

- estimate if a message has gone viral in some local group of users;
- estimate if a message has gone viral on a global scale.

If the first metric is pretty straightforward to be measured, in order to define and calculate the last two measures, we need to detect the so called "communities" from $G$ first. Once these clusters of nodes have been extracted, we can estimate if a tweet reached a relevant, although limited, popularity inside one or few communities, or if it reached a wider diffusion through the majority of the communities in our retweet network.

Throughout this paper, communities can be informally defined as non-overlapping sub-graphs of $G$ characterised by a high cohesion (i.e., many internal links) and high separation (i.e., few links connecting each other). Hence, a community detection task produces a partition $C = \{c_1, c_2, \ldots, c_n\}$ of $G$, so that $\forall i, j : c_i \cap c_j = \varnothing$, and $\bigcup_{c_i \in C} c_i = G$. Stated that the perfect partition of a graph is a computationally intractable problem, this informal definition of a community has been exploited to design many different clustering strategies based on efficient heuristics. In (26), we used the so called Louvain algorithm (32) because of its efficiency; in fact, the method is based on a modularity optimisation heuristic, and computational running time has order linear with the size of the network[2]. However, the most relevant aspect was to assign a higher level interpretation of those communities, following the common intuition that users with a strong political preference in Twitter tend to follow leaders and other ordinary users aligned with them. Then, we analysed the most common *bi-grams* in the users' tweets to identify a characterising stance towards immigration within each community, and whether such stance is mainly positive, negative or uncertain. The resulting communities and the respective stances can be found in Table 1.

The previous analysis highlights how the debate is strongly fragmented into communities that can be easily mapped to the multi-party system that traditionally characterises the recent Italian political landscape. Indeed, the groups found can all be traced back to different political parties and alliances, also because for each cluster was easy to find the highest in-degree nodes corresponding to leaders and influencers whose political orientation is clear, each one with its own stance towards immigration (26).

News media, either more politically oriented or neutral, are always central in the clusters; as detailed in Section 2.2, the segregated structure affects the diffusion of content, with the clusters linked to the opposition parties particularly affected by the echo chamber effect, such that some news are very popular and keep circulating inside those communities, and conversely other news rarely enters these self referential groups. Finally, some right-leaning, anti-immigration communities were more affected by the presence of news media and journalists that are well known for being misinformation spreaders.

## 2.2 Diffusion of external content on $G$

In our retweet network $G$ we study the diffusion of content - specifically, the diffusion of URLs referring to content external to Twitter, e.g., a web link to an article published by a news outlet. Indeed, every URL that is shared on the network will have one or more Original Posters (OPs), i.e., users that will inject it into the network through tweets, and other users who will *retweet* the OP's tweet, fostering the diffusion of the URL on the network.

Knowing the community to which each user belongs, we break down the diffusion chains and understand the sharing patterns of a URL between the different clusters. Let URLs = $\{\text{url}_1, \text{url}_2, \ldots, \text{url}_m\}$ be the set of all the URLs that have been shared by at least one user in TWITIMM. Then, we can quantify how *heterogeneous* the reach of a URL is, in terms of how many different communities it reaches, by counting the number of shares of each $\text{url}_i \in$ URLs by each community $c$ by means of $s_c(\text{url}_i)$, and then computing an entropy measure $H(\text{url}_i)$ of the array of shares (26), defined

---

[2]It should be observed that for the sake of simplicity, the directed graph was changed to undirected before executing the Louvain algorithm.

as

$$H(\text{url}_i) = -\sum_{c \in C} s_c(\text{url}_i) \ln(s_c(\text{url}_i)) \qquad (1)$$

that enables us to associate to each $\text{url}_i \in$ URLs a quantity that, whether it takes higher or lower values, tells us respectively how much an external content is able to spread across different clusters or instead how much it remains trapped in a "bubble"; such bubbles could then be identified as echo-chambers given the supposedly similar political stance of the users sharing the same external resource.

### 2.3 Untrustworthiness index

We now get to the heart of the present contribution by introducing the *Untrustworthiness* index $U$. Our aim is to discriminate between users that are more or less likely to engage with fabricated or manipulated contents online. With this goal in mind, we derived an *Untrustworthiness* index for each user that produced or retweeted at least one tweet containing an *url* to an external resource. This index measures how much they share or re-share news from unreliable sources.

The term *unreliable* here needs a solid definition. Online disinformation in Italy is often generated by pseudo-newspapers, allegedly satirical website (that never display a clear satirical intent in their content), blogs or unregistered media outlets (16). Such outlets have been targeted by independent fact checking organisations, creating blacklists that enlist those websites that publish partisan, unreliable or fabricated news. In this work we relied on the blacklists available from two main debunking sites in Italy, "*butac.net*" and "*bufale.net*": we obtained a selection of 24 websites that were consistent in publishing political mis- and dis-information, that were still active as of August 2019. Let's call $L^{\ominus}$ this list of blacklisted websites. As for the reliable media sources, we resorted to the Audiweb 2019 reports [3] to select a list of the top 100 information outlets by digital accesses; we then considered the web domains of every outlet in the list that were not blacklisted. This list of professional news outlets websites, no matter of their political orientation, are called $L^{\oplus}$.

Let $T_v$ be the total number of tweets produced by user $v$ containing an URL from either of the two lists; similarly, if we consider $T_v^{\ominus}$ and $T_v^{\oplus}$ as the number of tweets produced by user $v$ containing an URL from respectively $L^{\ominus}$ or $L^{\oplus}$, then $T_v = T_v^{\ominus} + T_v^{\oplus}$ will hold. We can easily calculate $R_v = T_v^{\ominus}/T_v$, namely the ratio of tweets produced by $v$ containing a URL from an unreliable source (i.e. a source blacklisted by fact checkers) over the total number of tweets containing a URL (from both reliable and unreliable sources). In order to assess an account's reliability not only in terms of this ratio, but also as a function of its activity, we define the **Untrustworthiness** of user $v$ as the harmonic mean of $R_v$ and $T_v$:

$$U_v = \left( \frac{\frac{\max_v(T_v)}{T_v} + \frac{1}{R_v}}{2} \right)^{-1} \qquad (2)$$

We can see in Figure 1) the distribution of $U$ on our dataset TWITIMM. This index allows us to easily discriminate users according to how much the information they share is reliable: users with higher $U$ that tweeted hundreds of times are likely to be stubborn spreaders of disinformation; users with low $U$ that tweeted news consistently can be instead classified at worst as occasional sharers of disinformation.

We plot the distribution of U in TWITIMM in Figure 1. We can see that the vast majority of the users have low U, but the right tail of the distribution decreases slowly, highlighting a core of high-U user at the very end of the distribution. The shape of the distribution gives us a first confirmation: given that we took into account a high number of high-credibility sources and that, in general, they generate way more traffic than low-credibility sources, we expected the numbers to be low and the majority of the users to have a low U. Still, the presence of several users in the right tail tells us that the phenomenon of sharing low-credibility content cannot be underestimated.

We tested for the robustness of $U$ on the TWITA dataset, checking the hypothesis that this score we assessed on our dataset, built on immigration-related words, could overestimate or underestimate the presence of low-credibility sources among the posts of the users. We computed the same score over the same period of one year, from August 2018 to August 2019 on a subset of users that can be found in both datasets (i.e., TWITIMM and TWITA). Results are consistent with those found in our dataset: 99% of the nodes have an overall $U$ in TWITA which is within a $\pm 0.1$ range from their own $U$ in TWITIMM.

---

[3]http://www.audiweb.it/

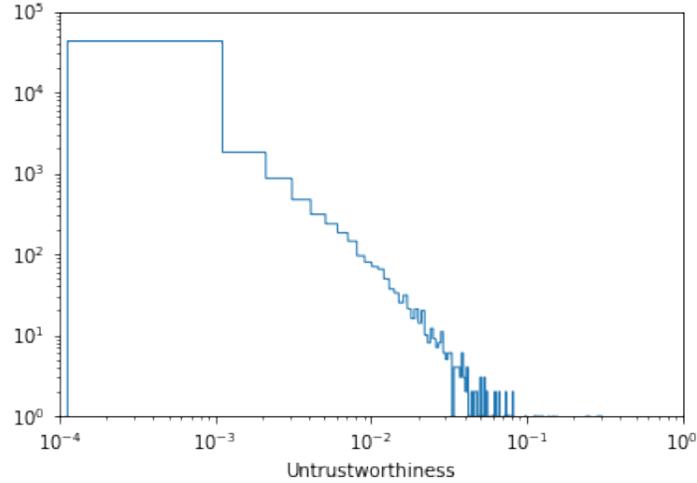

Figure 1: Distribution of the Untrustworthiness index of Twitter accounts in TWITIMM, i.e., accounts involved in the immigration debate. The numbers, weighted by the activity of the users, tend to be very low; in logarithmic scale we can see that the vast majority of the users have low U, but the right tail of the distribution decreases slowly, highlighting a core of high-U user at the very end of the distribution.

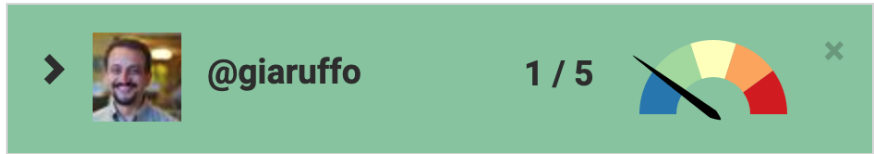

(a) One of the authors' Twitter account is likely human

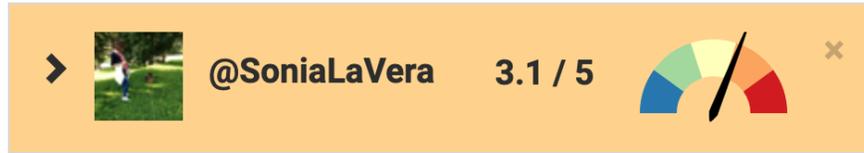

(b) The account with highest Untrusthworthiness ($U = 0.39$) in TWITIMM is likely bot

Figure 2: BotScore calculated on two different Twitter accounts (results from *https://botometer.osome.iu.edu*) using non-raw scores that span from 0 to 5.

## 2.4 BotScore

In order to quantify the impact of social bots on the Twitter debate around migrations, we first need to detect them. In our study we use the *Botometer* service (33), a tool developed by the researchers at OSoMe, the Observatory on Social Media at Indiana University. *Botometer* receives a Twitter user id as input, and it returns a set of scores that assess the "botness" of the corresponding account on multiple dimensions, leveraging a set of different classifiers trained to classify several types of bots, such as spammers, astroturfs or financial bots. The models used to execute the classification are trained using different feature sets that include network metrics, as well as text based attributes. Since these models have been inferred from corpora of tweets that were written mainly in English, for the purposes of our research, we had to leave out text based features from the classification task. Fortunately, *Botometer* allows the option of using the *overall raw score* (hence called BotScore), i.e. a score ranging in the interval $[0, 1]$ that assesses the likelihood of an account to be controlled by a bot by means of a language-independent classifier: see Figure 2 for two illustrative examples of calculating the BotScore on a couple of Twitter accounts.

In Fig. 3 we plot the distribution of the BotScore (BS) obtained by running Botometer on users in $N$. In the same figure, we also show a baseline distribution from a random sample of accounts of size equal to $|N|$. We extracted this sample of Italian Twitter users from the TWITA database (see Section 2.1), keeping the same temporal distribution of daily unique users found in TWITIMM. The two distributions are significantly different according to the Mann-Whitney test ($p \leq 10^{-4}$). Quite interestingly, the predominance of accounts likely controlled by humans over accounts likely controlled by bots is more pronounced among the tweets related to the immigration debate than among the randomly selected tweets related to different topics. Although we do not have any explanation for this, we suspect divisive topics are more engaging for real users with respect to. other conversations.

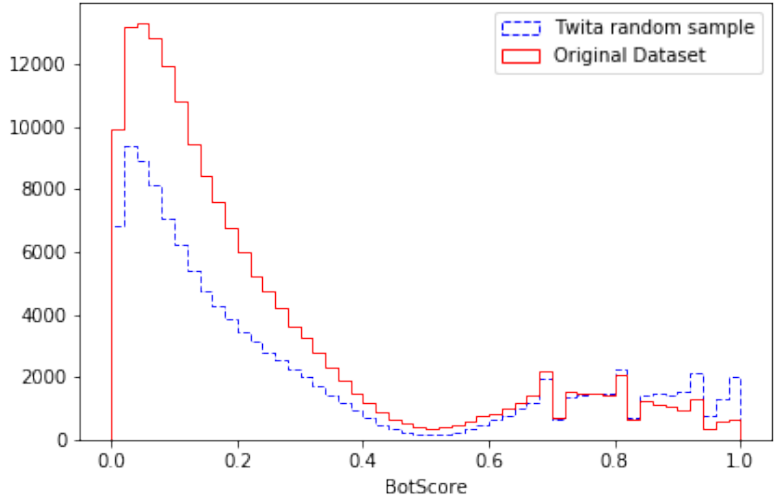

Figure 3: Distribution of botscores of Twitter accounts in TWITIMM, i.e., accounts involved in the immigration debate, compared to a random sample from Italian tweets produced in the same period, and equivalent size.

## 3 Results

Following the methods described in Section 2, we run a quantitative analysis on our set of nodes and communities to better understand which actors are involved in the debate and their communication dynamics. Hence, we characterise the users with respect to the following measures:

- Untrustworthiness U; this way we can gain insights on how much users from different communities tend to engage with news coming from sources identified as unreliable;
- BotScore BS, to understand how many automated accounts are distributed across the communities.

Finally, we will scrutinise the impact of U and BS on the diffusion of URLs on the retweet network $G$.

### 3.1 Relationships between Untrustworthiness and BotScore

We compute the Untrustworthiness of each user as described in Section 2.3. In Fig. 1 we plot the general distribution of U across the dataset, while in Fig. 4 we disaggregate it by the communities found in $G$ (described in Table 1) and, finally, we test the disaggregated distributions against a random reshuffling in Figure 5.

The interpretation of this result is meaningful and it strongly corroborates the qualitative characterisation of the clusters reviewed in Sec. 2.3; in fact the distribution of U in RT2 is strikingly different from the others. U scores in RT2 are indeed much higher, with a relevant number of users with high Untrustworthiness. Recall that RT2 is also the second largest community (see Table 1), already identified as an anti-immigration cluster, whose higher degree nodes correspond to accounts controlled by politicians, newspapers and celebrities who are publicly and vocally against foreign immigrants, often in close liaison with nationalist and right-wing parties. On the contrary RT1 seems to have very few users in the right tail of the distribution, i.e. with high U, even though it is by far the largest community. Mainly, the distributions seem to show a characteristic "untrustworthiness fingerprint" for each of the clusters, and this hypothesis holds against a randomisation of the community assignments (Fig.5). The differences between the distributions for each community and their random counterparts are all statistically significant (Mann-Whitney test, $p \leq 10^{-4}$).It is also possible to observe that RT2's original data shows that BS scores are constantly higher than the expectation from a random reshuffle, while RT1 shows an opposite behaviour.

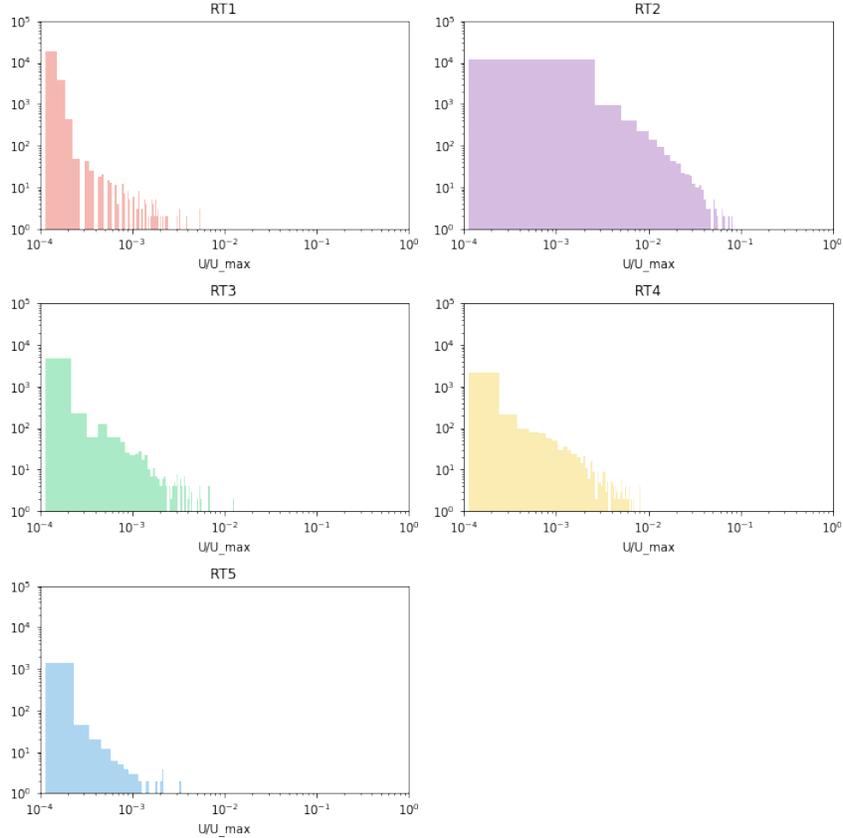

Figure 4: $U$ among communities. RT2, previously identified as an anti-immigration cluster, clearly exceeds the average number of high-U users by far, showing a higher and longer right tail than the other communities, suggesting that the prevalence of low credibility content in this cluster is much higher.

On the same note we compute the BotScore, as described in Section 2.4, and we plot the BS distributions community-wise (Fig.6). The BS distribution are characterised differently with respect to U scores. We are not immediately able to identify one or two communities that stand out among the others in terms of a much higher presence of bots. Nonetheless, we can still say that there are clear differences between the clusters. As for the case of the Untrustworthiness, the distributions are statistically significantly different from a random baseline (Mann-Whitney test, $p \leq 10^{-4}$). In all the communities, we can observe evidence that accounts are controlled more likely by humans than by bots (the histograms' bins are always much more pronounced for BS $< 0.5$). Nevertheless, it looks like that the ratio of accounts likely controlled by bots and accounts likely controlled by humans (fixing a threshold to BS $= 0.5$ as a rule of thumb) is lower for RT1 and higher for RT3 and RT5.

### 3.2 Diffusion of URLs and impact of the Original Posters

At this point, U and BS scores do not seem to be strongly related, even though they show their own peculiarities in terms of how they are distributed across the different communities. It is interesting though to uncover the role these users' features play in the diffusion of URLs on the network. To do so, we consider a set of $\approx 700$ URLs, namely all the URLs in our dataset that were shared more than 100 times. These URLs spread all across the network; the extent of their diffusion is quantified not only by the number of retweets but also through an entropy measure, as described in Section 2.2, that informs us on the heterogeneity of the reach of the URL in terms of number of different communities retweeting it. Therefore, we are able to characterise each URL on different dimensions: entropy $H$, the number of retweets and the features (BotScore, Untrustworthiness) of the users that are sharing it. In Fig. 7 we cross-check these dimensions in order to evaluate the interplay between U and BS scores, as well as their impact on the URLs diffusion. For each URL we computed the average U and BS scores for all users sharing it. As we can see, there is a clear shift in the Untrustworthiness as the BotScore rises. A cloud of red dots, all located above BS $\gtrsim 0.25$, tells us that - on average - those URLs retweeted by users with higher BotScore are often retweeted by users with high

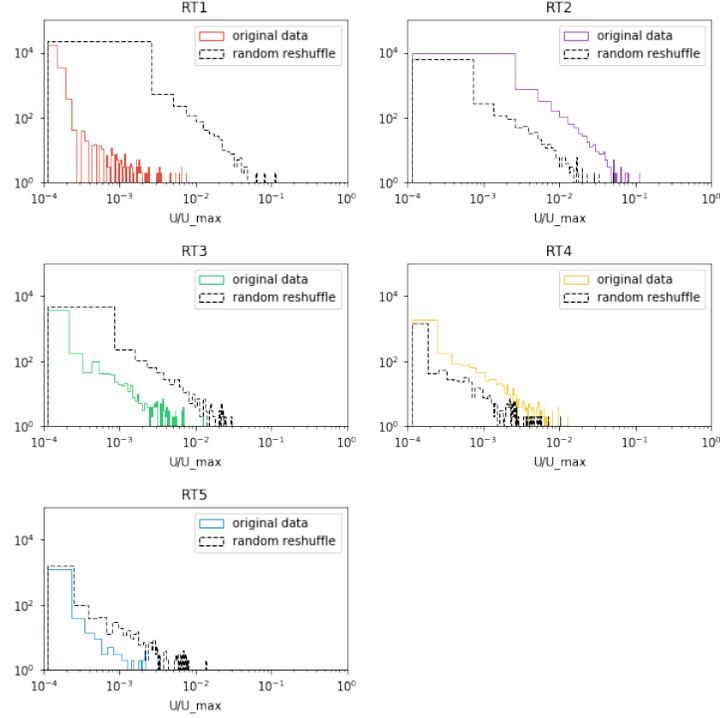

Figure 5: Disaggregated distribution of $U$ tested against a randomly reshuffled community partitioning. All the distributions are statistically non-random according to the Mann-Whitney test ($p \leq 10^{-4}$)

Untrustworthiness, suggesting an interesting correlation between these two dimensions. Also entropy comes into play: the highest number of dark red dots (high-U URLs) can be found in the low-entropy area of the plot.

We want to investigate the dynamics of diffusion by further exploring the interplay between U and BS. We focus on the role of the Original Posters (OPs), i.e. those that first *tweet* an URL which is then retweeted by others. Similarly to Fig. 7, in Figure 8 we consider the entropy, the average BotScore and the average Untrustworthiness, this time for all the URLs whose OPs have an high BotScore ($BS > 0.75$). We can clearly identify a set of URLs with low entropy, high average BS and high average U: the URLs injected in the network by the alleged bots seem to be very low credibility content that do not reach a great variety of communities. In light of this, we can argue that those OPs that are likely bots foster the echo-chamber effect that allows some low credibility content to reach a high number of shares while remaining confined in a small number of communities.

### 3.3 Probability of success

We analyse now the impact of the Original Posters on the diffusion of content also by quantifying the probability of success of a URL given the BotScore and Untrustworthiness of its OPs. Indeed, one URL can have more than one OP - namely all those users who tweeted the URL and are the seeds of different independent retweet cascades. In such cases we consider the average BS and U of all the OPs for each URL. Furthermore, we says that a URL is successful if, given the distribution of the number of retweets per URL in our dataset, it falls in the fourth quartile, i.e. it is among the top 25% of the most retweeted URLs.

On these premises, we resort to Bayes theorem to compute the conditional probability for an URL of being successful if the average BS of its OPs is above a certain value *x*:

$$\text{P(RT} \geq \text{t} \mid \text{avg. BS} \geq \text{x)} = \frac{\text{P(avg. BS} \geq \text{x} \mid \text{RT} \geq \text{t}) \cdot \text{P(RT} \geq \text{t})}{\text{P(avg. BS} \geq \text{x})} \tag{3}$$

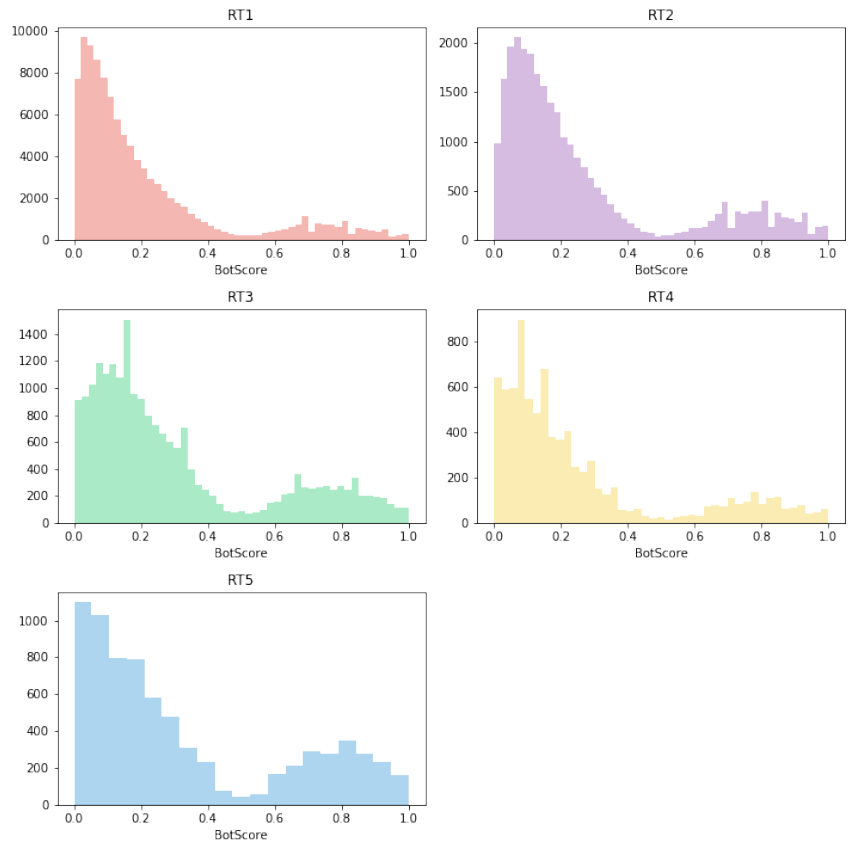

Figure 6: BS distribution disaggregated by community. Here a semi-log axis scale is used to better highlight the differences between the clusters. We can see that the right tail of the distribution - namely, all individuals with high BotScore - is generally lower than the left tail. This is particularly true for RT1.

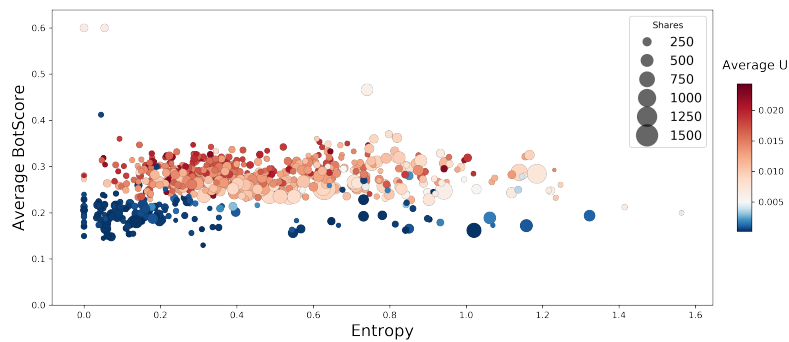

Figure 7: BotScore as a function of the entropy: each dot is a URL. BotScore and Untrustworthiness by URL are computed as the average values of the scores of the users retweeting each URL.

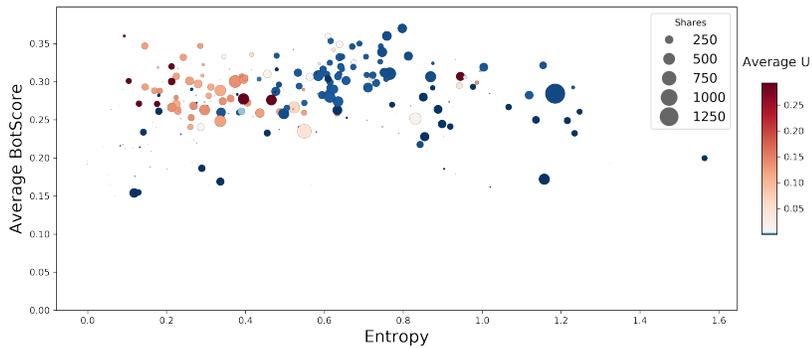

Figure 8: Relationship between entropy, Untrustworthiness and BotScore for URLs originally shared by OPs with high BS (BS > 0.70). We can clearly identify a set of URLs with low entropy, high average BS and high average U: the URLs injected in the network by the alleged bots seem to be very low credibility content that do not reach a great variety of communities, fostering an echo-chamber effect.

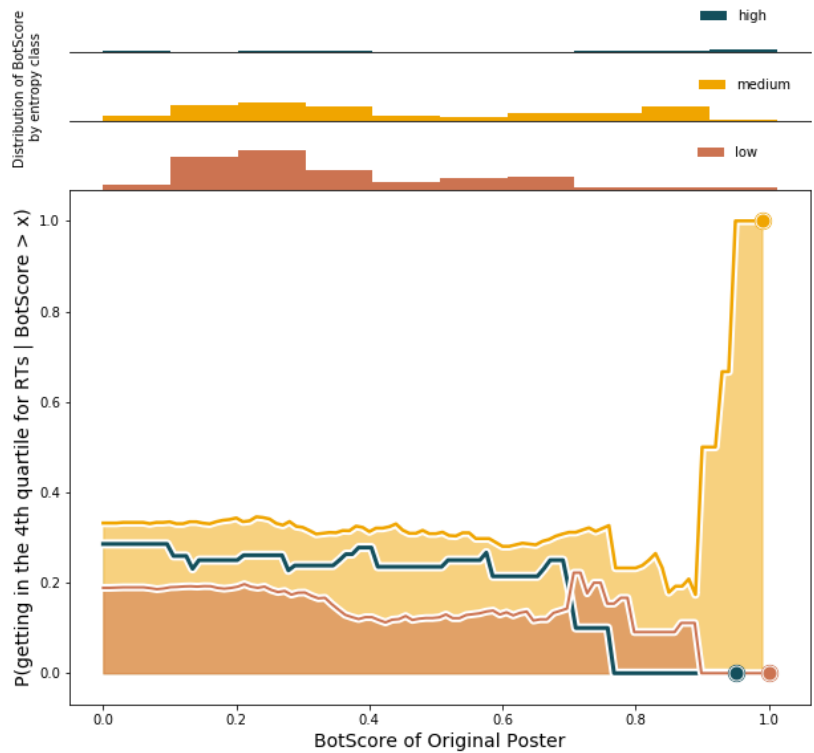

Figure 9: Probability of success for a URL given the BotScore of the OP. Medium entropy URLs tend in general to be more successful than the others, regardless of the BS of the original posters

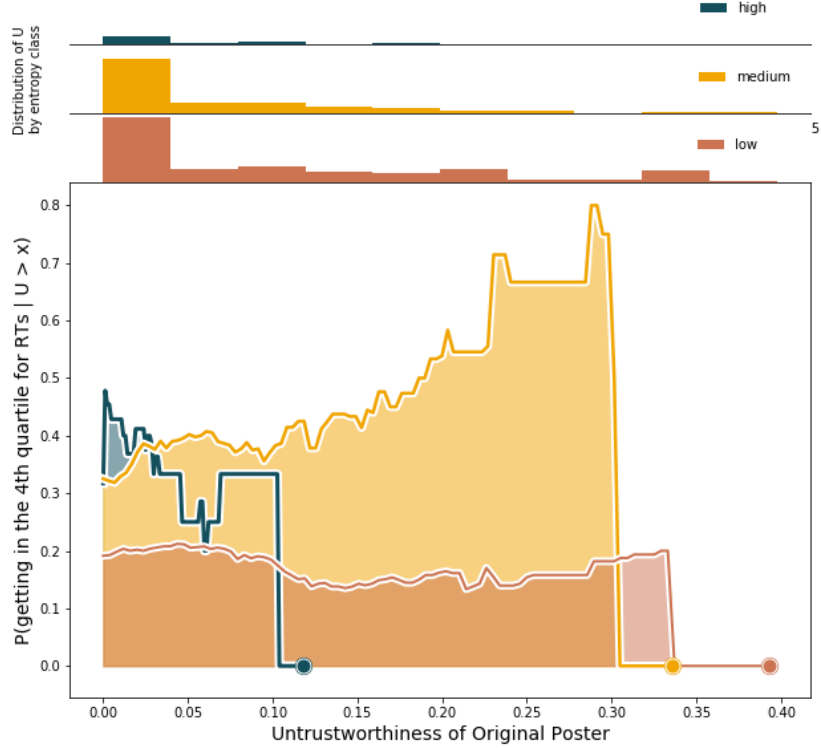

Figure 10: Probability of success for a tweet given the Untrustworthiness of the OP. As seen in Fig. 9 for the BotScore, medium entropy URLs tend to dominate over the others, always obtaining more success, with the only exception of the high entropy and low-U URLs, that probably represent a core of reliable, mainstream media content that are widely diffused through the network. Evidence in this Figure and in Fig. 9 suggest that the key for an URL to be successful is to be diffused beyond its originating community, but without getting too far from it - or, better, without being shared in too many other communities.

.

where *RT* stands for the number of retweets of an URL and *t* is the fixed threshold number of retweets corresponding to the fourth quartile. The same applies to the success probability conditioned by the average OP Untrustworthiness:

$$P(RT \geq t \mid \text{avg. U} \geq x) = \frac{P(\text{avg. U} \geq x \mid RT \geq t) \cdot P(RT \geq t)}{P(\text{avg. U} \geq x)} \quad (4)$$

In Figure 9 and 10 we can see how the probabilities, defined respectively in Eq.3 and 4, vary as a function of the threshold variable $x$, with fixed $t$. Both plots display the probabilities for URLs having low ($\leq 0.4$), medium ($\leq 0.9$) and high ($> 0.9$) entropy. The upper part of both figures shows the distribution of the feature under examination (BS and U) by entropy class.

We can gather interesting insights. By checking the distributions of BS and U by entropy class, we see that the high entropy category is, in general, scarcely populated. This has practical implications on the high-entropy probability, since there are fewer data points we can use to calculate it; this is particularly evident in Fig.10, where the complete absence of URLs with high entropy and average OP Untrustworthiness $\gtrsim 0.10$ abruptly brings the probability to 0.

The entropy class clearly seems to be discriminating between different levels of success. In both cases, medium entropy URLs are those getting the most retweets: they are gaining more success, making the *medium* entropy - namely, the diffusion of an URL across different communities, but *not too many* - a key factor in the success. Entropy overshadows the effect of $U$ and $BS$ over the eventual success of an URL since, for each entropy class, the probabilities remain essentially constant for any given value of BS and U, suggesting that they do not play a key role in defining the "faith" of the diffusion of the URL: there are likely other factors affecting such dynamics, and entropy seems to be definitely one of them.

## 4  Discussion

In this work we conducted an empirical analysis to evaluate the presence and popularity of low credibility media content in the Italian Twitter debate on immigration. We carried out an extensive analysis on more than 6M tweets, quantifying the presence of bots among the Twitter users and characterising the users with an *Untrustworthiness* score $U$, i.e. a measure that allows us to discriminate between those accounts that frequently engage with untrustworthy media content and those who do not. This characterisation is helpful to understand whether there are communities of users that tend to be more exposed to disinformation, also in relation with their political leaning and their stance towards immigration.

We found $U$ to be insightful in this specific case study. By mapping it on the community structure, we notice that there are peculiar trends for the different clusters, suggesting that some groups are indeed more affected by the consumption of low credibility content than others. This is particularly true for community *RT2*, that was previously identified as a community with negative stance towards immigration, centred around media outlets that could be defined as *ambiguous* at the very least, if not openly disinformation spreaders. We cannot see similar patterns for the presence of automated accounts; in fact, after using Botometer (33) to assess the distribution of bots among different clusters, we found statistical evidence of non-randomness in the overall dataset when compared to a random, topic-neutral data sample, suggesting a stronger prevalence of bots in this specific debate; when it comes to the cluster analysis, three communities - *RT2, RT3* and *RT5* - display a slightly higher right tail, showing a higher presence of bots in these communities.

This analysis relies on the good performance of the language-independent model used for the classification; most importantly, it does not necessarily imply a malicious nature of the automated accounts. In order to estimate the level of engagement of bots with disinformation content, we cross-checked the BotScore (BS) with $U$ by analysing the most retweeted URLs and the users involved. We see in Fig. 7 that the BS of the users acts as a good discriminating feature, as we note that for BS $\gtrsim 0.25$ the URLs are shared by users with high $U$, especially for the low entropy URLs that are not retweeted by many different clusters, but rather remain confined to a selected few. These URLs seem to be those where these two aspects - action of bots and engagement with disinformation - strongly come together.

Furthermore, we were particularly interested in exploring the role of the Original Posters (OPs), to understand if bots play a relevant role in *injecting* content in the network. Indeed, several of those URLs being shared first by alleged bots are then retweeted almost exclusively by users with very high $U$, showing a clear interplay between these two features. These URLs have a very low entropy, suggesting that bots do play a role in introducing content that bounces always within the same walls, obtaining several shares but in a small number of communities. On the other hand, in order to allow bot-shared content to reach a high number of different communities, this content must pass through low-U users.

Finally, we look into the factors that might affect the success of a piece of media content. Having defined *success* as the condition of being among the top 25% most retweeted URLs, we computed the conditional probability for an URL to fall in this court given $U$ and BS of its OPs. Unexpectedly, we see that neither $BS$ nor $U$ seem to be really decisive in determining the success of an URL: the probabilities follow very similar trends (Figures 9 and 10). Entropy instead stands out: for both probabilities the class of medium entropy URLs emerges clearly, keeping a high probability throughout the whole range of thresholds set for BS and $U$, and completely dominating on the high values. This is a relevant results: according to our data, the key for an URL to be successful is to be diffused beyond its originating community, but without getting too far from it - or, better, without being shared in too many other communities.

We believe that the present work has some limitations to acknowledge and some talking points to address. Particularly:

- being an empirical work, the results of the analysis strongly depend on the data. Having selected a very general, neutral set of keywords to filter the tweets, we believe that our dataset is highly representative of the debate around immigration in Italy, as it has also been discussed in (26). Still, Twitter Stream APIs come with some constraints that could, in principle, limit the representativity of the dataset; moreover, we must remember that it is not straightforward to generalise any conclusion drawn from a sample of users from a single social media to the entire population;
- the Untrustworthiness index $U$ is based on the selection of reliable and unreliable information outlets. Therefore, special care should be taken when selecting the sources that allow us to label them;
- on the same note, we would like to emphasise again how the phenomenon of disinformation is extremely complex and multi-faceted. In the present work we opted for a binary characterisation of the media outlets.

This characterisation is based upon the *intentionality* of the disinformation action: the unreliable, blacklisted outlets are those that are *deliberately* producing and spreading fake or ambiguous content, as recognised by third-parties disinformation watch-lists. Nonetheless, there are many other aspects of disinformation other than the simple deliberateness of the action. Ambiguous content can also be generated or amplified by well-known and trusted news media, often unintentionally: this is a very different scenario, that requires a different approach, likely focused on the individual news stories rather than on the media outlet. With this work we decided to cover the other end of the disinformation spectrum, focusing on the intent and processes of the publishers, as we extensively argued in Section 1.

## 5   Conclusions

Overall, our results enable us to answer the questions formulated in Section 1. By crossing information on the users' habitual consumption of low credibility media content with a statistical analysis on the presence of bots, we were able to observe the interplay between the two. Even though the mere presence - the distributions across the communities - of automated accounts and of users with high $U$ do not show blatant signs of correlation, they clearly interact when it comes to the diffusion of content in segregated communities: bots seem to be particularly involved in the sharing of untrustworthy content and play a role in injecting such content in the network, with the said content obtaining a high number of shares while remaining confined in few communities, fostering an echo-chamber effect. A high-level study of the global dynamics of content success does not tell us that Untrustworthiness and BotScore are decisive in the virality; nonetheless, a detailed reconstruction of the actual retweet cascades could be definitely helpful to have a more precise idea of the role of bots in the spreading of disinformation, expanding the insights gathered in our experiment. While we leave this as future work, in light of our results we are able to state that engagement with disinformation media is not equally distributed across the network of Twitter users and that automated accounts are involved in the spreading of unreliable content, particularly among those users who interact more often with such content, though without being crucial for its virality.

## References

[1] Merriam-Webster: The Real Story of 'Fake News,'. Accessed: 04-Jun-2021. https://www.merriam-webster.com/words-at-play/the-real-story-of-fake-news

[2] Wardle, C., Derakhshan, H.: Information disorder: Toward an interdisciplinary framework for research and policy making. Council of Europe report **27**, 1–107 (2017)

[3] Economist, T.: Italy's populist right looks menacing. Accessed: 04-Jun-2021. https://www.economist.com/europe/2021/05/22/italys-populist-right-looks-menacing

[4] Vargo, C.J., Guo, L., Amazeen, M.A.: The agenda-setting power of fake news: A big data analysis of the online media landscape from 2014 to 2016. New Media & Society **20**, 2028–2049 (2018)

[5] Allcott, H., Gentzkow, M., Yu, C.: Trends in the diffusion of misinformation on social media. Research & Politics **6**, 205316801984855 (2019). doi:https://doi.org/10.1177/2053168019848554

[6] Guess, A., Nyhan, B., Reifler, J.: Selective exposure to misinformation: Evidence from the consumption of fake news during the 2016 us presidential campaign. European Research Council **9**(3), 4 (2018)

[7] Tacchini, E., Ballarin, G., Vedova, M.L.D., Moret, S., de Alfaro, L.: Some like it hoax: Automated fake news detection in social networks. CoRR **abs/1704.07506** (2017). 1704.07506

[8] Lazer, D.M.J., Baum, M.A., Benkler, Y., Berinsky, A.J., Greenhill, K.M., Menczer, F., Metzger, M.J., Nyhan, B., Pennycook, G., Rothschild, D., Schudson, M., Sloman, S.A., Sunstein, C.R., Thorson, E.A., Watts, D.J., Zittrain, J.L.: The science of fake news. Science **359**(6380), 1094–1096 (2018). doi:https://doi.org/10.1126/science.aao2998. https://science.sciencemag.org/content/359/6380/1094.full.pdf

[9] Bodrunova, S.S., Litvinenko, A.A.: New media and political protest: The formation of a public counter-sphere in russia, 2008-12. In: Russia's Changing Economic and Political Regimes: The Putin Years and Afterwards, pp. 29–65 (2013)

[10] de Saint Laurent, C., Glaveanu, V., Chaudet, C.: Malevolent creativity and social media: Creating anti-immigration communities on twitter. Creativity Research Journal **32**(1), 66–80 (2020). doi:https://doi.org/10.1080/10400419.2020.1712164. https://doi.org/10.1080/10400419.2020.1712164

[11] Bodrunova, S.S., Litvinenko, A.A., Gavra, D.P., Yakunin, A.V.: Twitter-based discourse on migrants in russia: the case of 2013 bashings in biryulyovo. International Review of Management and Marketing **5**(1S) (2015)

[12] Siapera, E., Boudourides, M., Lenis, S., Suiter, J.: Refugees and network publics on twitter: Networked framing, affect, and capture. Social Media+ Society **4**(1), 2056305118764437 (2018)

[13] Humprecht, E.: Where 'fake news' flourishes: a comparison across four western democracies. Information, Communication & Society **22**(13), 1973–1988 (2019)

[14] Chenzi, V.: Fake news, social media and xenophobia in south africa. African Identities, 1–20 (2020)

[15] Gualda, E., Rebollo, C.: The refugee crisis on twitter: A diversity of discourses at a european crossroads. Journal of Spatial and Organizational Dynamics **4**(3), 199–212 (2016)

[16] Pierri, F., Artoni, A., Ceri, S.: Investigating italian disinformation spreading on twitter in the context of 2019 european elections. PLOS ONE **15**(1) (2020)

[17] Shin, J., Jian, L., Driscoll, K., Bar, F.: The diffusion of misinformation on social media: Temporal pattern, message, and source. Computers in Human Behavior **83**, 278–287 (2018). doi:https://doi.org/10.1016/j.chb.2018.02.008

[18] Ferrara, E., Varol, O., Davis, C., Menczer, F., Flammini, A.: The rise of social bots. Communications of the ACM **59**(7), 96–104 (2016)

[19] Vosoughi, S., Roy, D., Aral, S.: The spread of true and false news online. Science **359**(6380), 1146–1151 (2018)

[20] Shao, C., Ciampaglia, G.L., Varol, O., Yang, K.-C., Flammini, A., Menczer, F.: The spread of low-credibility content by social bots. Nature communications **9**(1), 1–9 (2018)

[21] Stella, M., Ferrara, E., De Domenico, M.: Bots increase exposure to negative and inflammatory content in online social systems. Proceedings of the National Academy of Sciences **115**(49), 12435–12440 (2018)

[22] Bessi, A., Ferrara, E.: Social bots distort the 2016 us presidential election online discussion. First Monday **21**(11-7) (2016)

[23] Suárez-Serrato, P., Roberts, M.E., Davis, C.A., Menczer, F.: On the influence of social bots in online protests. preliminary findings of a mexican case study. CoRR **abs/1609.08239** (2016). 1609.08239

[24] Forelle, M., Howard, P.N., Monroy-Hernández, A., Savage, S.: Political bots and the manipulation of public opinion in venezuela. CoRR **abs/1507.07109** (2015). 1507.07109

[25] Abokhodair, N., Yoo, D., McDonald, D.W.: Dissecting a social botnet: Growth, content and influence in twitter. CoRR **abs/1604.03627** (2016). 1604.03627

[26] Vilella, S., Lai, M., Paolotti, D., Ruffo, G.: Immigration as a divisive topic: Clusters and content diffusion in the italian twitter debate. Future Internet **12**(10), 173 (2020)

[27] Basile, V., Lai, M., Sanguinetti, M.: Long-term social media data collection at the university of turin. In: Fifth Italian Conference on Computational Linguistics (CLiC-it 2018), pp. 1–6 (2018). CEUR-WS

[28] Guille, A., Hacid, H., Favre, C., Zighed, D.A.: Information diffusion in online social networks: A survey. SIGMOD Rec. **42**(2), 17–28 (2013). doi:https://doi.org/10.1145/2503792.2503797

[29] Conover, M., Ratkiewicz, J., Francisco, M., Gonçalves, B., Menczer, F., Flammini, A.: Political polarization on twitter. In: Proceedings of the International AAAI Conference on Web and Social Media, vol. 5 (2011)

[30] Feller, A., Kuhnert, M., Sprenger, T.O., Welpe, I.M.: Divided they tweet: The network structure of political microbloggers and discussion topics. In: Fifth International AAAI Conference on Weblogs and Social Media (2011)

[31] Lai, M., Tambuscio, M., Patti, V., Ruffo, G., Rosso, P.: Stance polarity in political debates: A diachronic perspective of network homophily and conversations on twitter. Data & Knowledge Engineering, 101738 (2019)

[32] Blondel, V.D., Guillaume, J.-L., Lambiotte, R., Lefebvre, E.: Fast unfolding of communities in large networks. Journal of statistical mechanics: theory and experiment **2008**(10), 10008 (2008)

[33] Sayyadiharikandeh, M., Varol, O., Yang, K.-C., Flammini, A., Menczer, F.: Detection of novel social bots by ensembles of specialized classifiers. In: Proceedings of the 29th ACM International Conference on Information & Knowledge Management, pp. 2725–2732 (2020)